\documentclass[graybox]{svmult}
\usepackage{mathptmx}       
\usepackage{helvet}         
\usepackage{courier}        
\usepackage{type1cm}        
\usepackage{makeidx}         
\usepackage{graphicx}        
\usepackage{multicol}        
\usepackage[bottom]{footmisc}

\newcommand{\Msun}{M$_{\odot}$}
\newcommand{\J}{M$_{\rm JUP}$}
\newcommand{\apjl}{ApJL}%
          
 \newcommand{\apj}{ApJ}%
          
\newcommand{\aap}{A\&A}%
\newcommand{\nat}{Nature}
\newcommand{\arcsec}{"}

\newcommand{\mnras}{MNRAS}%

\begin{document}

\title{Searching for Brown Dwarf Outflows}
\author{E.T. Whelan, T.P. Ray, F. Bacciotti, S. Randich, A. Natta}
\institute{E.T. Whelan, T.P. Ray \at Dublin Institute for Advanced Studies, \email{ewhelan@cp.dias.ie}
\and F. Bacciotti, S. Randich, A. Natte \at Osservatorio Astrofisico di Arcetri}
%
%
\maketitle

\abstract{As outflow activity in low mass protostars is strongly connected to accretion it is reasonable to expect accreting \index{brown dwarf} brown dwarfs to also be driving outflows. In the last three years we have searched for brown dwarf outflows using high quality optical spectra obtained with UVES on the VLT and the technique of \index{spectro-astrometry} spectro-astrometry. To date five brown dwarf outflows have been discovered. Here the method is discussed and the results to date outlined.}

\section{Introduction}

It is now apparent that protostellar-like outflows commonly accompany the formation and evolution of brown dwarfs (BDs; 12, 16, 18). The overall motivation of this project is to investigate the validity of the accretion/ejection models for the formation 
of solar-mass stars at the lowest masses, through the study of BD outflows. Ultimately, results will be used to form a better understanding of the outflow mechanism in general. Much of 
what is known about low mass star formation and in 
particular the connection between magntospheric accretion and outflow activity comes from studying the classical T Tauri stars (CTTS). Forbidden emission lines (FELs) \index{forbidden emission lines}
have proved to be effective tracers of outflow activity in CTTSs and to date they have been used to explore the kinematics, morphology and physical conditions of jets, at high angular resolution (13). The so-called traditional tracers of CTT jets, i.e. 
[OI]$\lambda\lambda$6300,6363, [SII]$\lambda\lambda$6716,6731, [NII]$\lambda$6583 lines, are found 
in the spectra of BDs. This finding was the first indication that BDs launch outflows.

BDs outflows are difficult to detect as they are faint and FEL tracers are only extended on milli-arcsecond scales. Hence long exposure times ($\sim$ 1.5 hours with the VLT) and high angular resolution techniques are needed. Our approach is to obtain high spatial resolution spectra with the UV-visual echelle spectrometer (UVES) on the VLT and to analyse the origin of key lines with spectro-astrometry. Table 1.1 lists the objects found to date to be driving optical outflows. Four of these objects have derived masses placing them well within the BD mass range. The fifth, ISO-ChaI 217 has an estimated mass of 80 \J\ placing it just above the hydrogen burning mass limit (HBML).

\section{Targets, Observations and Analysis}

The BDs \index{brown dwarf} targeted for this study had all shown evidence of strong T Tauri-like accretion and several were already known to exhibit forbidden emission. High resolution UVES spectra were obtained, reduced using standard IRAF routines and, as stated above, analysed with spectro-astrometry. For ISO-Cha I 217 and ISO-Oph 32, spectra were obtained at orthogonal slit position angles (PA), allowing the PAs of the outflows to be estimated. For the majority of protostars with jets the FEL regions of their spectra are very obviously spatially extended, particularly at high velocities. For example the FELs of the CTTS DG Tau trace a $\sim$ 12\arcsec\ knotty jet (15).
However, when the emission is very compact and close to the source (as in the case of BDs) specialised methods such as spectro-astrometry have to be used to probe its nature. 
Spectro-astrometry utilises simple Gaussian Fitting to investigate the 
 offset, with respect to the star/BD, of emission line features smaller than the seeing disc of the observation.  
The result of this technique is an offset-velocity diagram, with the displacement of a spectral feature (e.g. the [OI]$\lambda$6300) line shown as a function of velocity and relative to the continuum centroid of the object (19). The beauty of this technique lies in the fact that its accuracy depends primarily on the signal to noise (S/N) of the observation. Formally, the spectro-astrometric accuracy is given as follows, $\sigma$ = ({\it \rm Seeing}) / (2.3548$\sqrt{N_{p}}$), where {\it N$_{p}$} is the number of detected photons.  Hence, even under conditions of poor seeing, a high accuracy can be achieved. Refer to Whelan et al. 2007 and Whelan $\&$ Garcia 2008, for further details on the spectro-astrometric technique, including information on spectro-astrometric artifacts and on how the S/N can be increased in BD spectra.

\begin{table}
\begin{tabular}{lllll}       
 \hline 
 Source                                   &RA (J2000)   &Dec (J2000) & Spectral Type &Mass (\Msun)     
 \\ 
\hline
ISO-ChaI 217                  &11 09 52.0                  &-76 39 12  &M6.2                 &0.08 [8]             
\\
2MASS1207-3932          &12 07 33.4                  &-39 32 54.0  &M8.5            &0.024 [10]
\\
 ISO-Oph 32                &16 26 22.05  &-24 44 37.5 &M8 &0.04 [11]     
 \\
 ISO-Oph 102                 &16 27 06.5     &-24 41 47.1 &M6.5  &0.06  [11]   
 \\
 LS-RCrA 1   &19 01 33.7  &-37 00 30 &M6.5 &0.04 [1]
\\  
 \hline  
\end{tabular}
\caption{Brown Dwarfs and a Very Low Mass Star Discovered to have Optical Outflows. The papers referencing the mass of these objects are given in square brackets in column 4.}

\label{tab1}
\end{table}

\begin{center}
\begin{figure}
\includegraphics[width = 9.8cm]{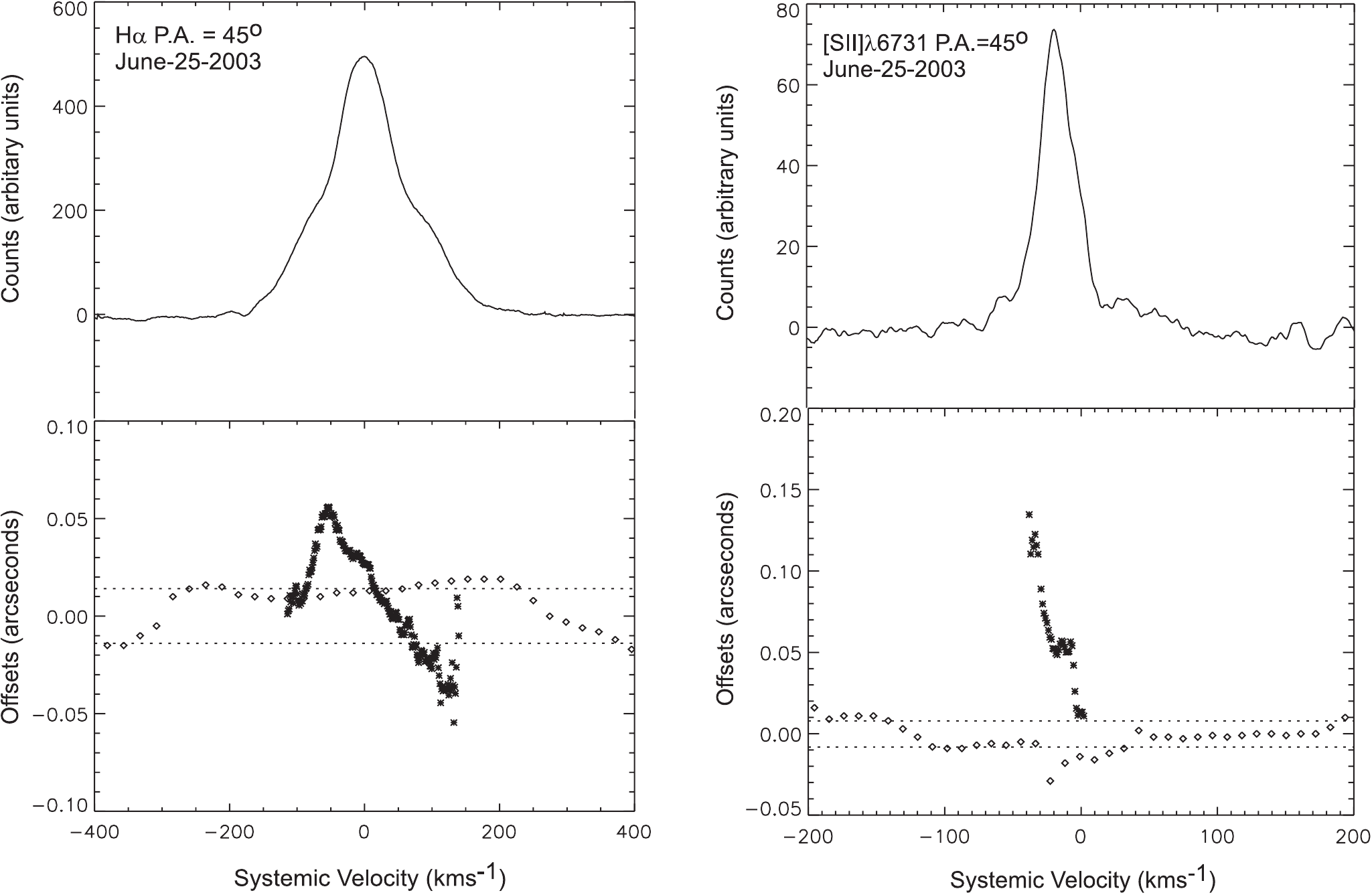}
\caption{The spectro-astrometric analysis of the H$\alpha$ and [SII]$\lambda$6731 lines in the spectrum of LS-RCrA 1. The black stars represent the offsets in the line, in the continuum subtracted spectrum, and the open diamonds the position of the continuum. The line and continuum were binned so that the S/N and thus the spectro-astrometric accuracy in both was comparable.  The dashed line represents the 1-$\sigma$ error in the spectro-astrometric analysis.} 
\end{figure}
\end{center}

\section{Results}

In this section we will briefly discuss the objects analysed to date (in order of their investigation) and the properties of the outflows uncovered. 

\begin{itemize}

\item{{\bf ISO-Oph 102 and ISO-Oph 32:} ISO-Oph 102 was the first BD investigated using spectro-astrometry and the results of this investigation were published in Whelan et al 2005.  Hence it is the first BD discovered to have an outflow. The [OI]$\lambda$6300, 6363, [NII]$\lambda$6583 and [SII]$\lambda$6731 lines were all detected along with the H$\alpha$ line which exhibited a P-Cygni like profile. Our investigation revealed the [OI] and [SII] lines to be offset to a distance of $\sim$ 100 mas at a velocity of $\sim$ -40 kms$^{-1}$. The [NII]$\lambda$6583 was too faint for spectro-astrometric analysis. Only blue-shifted emission was detected, as is commonly seen in CTTSs. The circumstellar disk is assumed to be obscuring the red-shifted part of the flow. The scale of the blueshifted offset would suggest a minimum 
(projected) disk radius of 100 mas ($\leq$ 15 AU at the distance 
of the $\rho$-Ophiuchi cloud) in order to hide the redshifted component. ISO-Oph 32 is a 40 \J\ BD which again was known to be a strong accretor (13). Our recent analysis revealed weak [OI]$\lambda$6300, 6363 emission, similar in strength to what we detected for 2MASS1207-3932. The [OI]$\lambda$6300 line is blue-shifted to a velocity of $\sim$ -30 kms$^{-1}$ and offset to $\sim$ 60 mas at 0$^{\circ}$ and $\sim$110 mas at 90$^{\circ}$. 
This constrains the PA of this outflow at $\sim$ 60$^{\circ}$ (E of N). 

}

\item{{\bf 2MASS1207-3932:} UVES spectra of the 24 \J\ BD 2MASS1207-3932 obtained in May 2006 revealed the presence of strong H$\alpha$ emission and the [OI]$\lambda\lambda$6300, 6363 emission lines. Only the [OI]$\lambda$6300 was strong enough for spectro-astrometric analysis. The line profile was double peaked with blue and redshifted emission at $\sim$ -8 kms$^{-1}$ and 4 kms$^{-1}$, respectively. The blue and red-shifted emission was found to be offset in opposite directions to $\sim$ 85 mas, revealing the presence of a bipolar optical outflow. 2MASS1207-3932 is now the lowest mass galactic object known to drive an outflow. The low radial velocities measured in the [OI]$\lambda$6300 line agree with the hypothesis that 2MASS1207-3932 has a near edge-on accretion disk (19).}

\item {{\bf ISO-Cha I 217:} Spectra of ISO-Cha 1 217 at orthogonal positon angles (0$^{\circ}$, 90$^{\circ}$) were obtained in September 2007. The FELs of [OI]$\lambda\lambda$6300, 6363, [SII]$\lambda\lambda$6716, 6731 and [NII]$\lambda$6583 were found to be bright, and spectro-astrometric analysis revealed the presence of a bipolar outflow. Systemic velocities of $\sim$ -20, +30 kms$^{-1}$ were measured at both 0$^{\circ}$ and 90$^{\circ}$ and, interestingly, the red-shifted emission was found to be twice as bright as the blue-shifted. This type of asymmetry has been seen before in CTTSs. Spectro-astrometry revealed the red and blue components to be offset to $\sim$ +/- 200 mas at 0$^{\circ}$ and +/- 50 mas at 90$^{\circ}$. This suggests a P.A. for the outflow from ISO-Cha I 217 of $\sim$ 15$^{\circ}$ (E of N). }

\item{{\bf LS-RCrA 1:} LS-RCrA 1 was the first BD shown to have forbidden emission (2). Early attempts to spatially map the FELs, using spectra taken with Magellan Inamori Kyocera Echelle ( MIKE ) on the Magellan II 
telescope failed due to the poor quality of the spectra and hence the faintness of the continuum emission (17). For this study UVES spectra, taken in June 2003, were obtained from the ESO data archive. Again bright [OI]$\lambda\lambda$6300, 6363, [SII]$\lambda\lambda$6716, 6731 and [NII]$\lambda$6583 are present in the spectrum and offsets of up to $\sim$150 mas were recovered. The H$\alpha$ line profile has blue and red-shifted ``humps" which our spectro-astrometric analysis has revealed to be offset in opposite directions. Hence, the H$\alpha$ line has a component originating in the outflow from LS-RCrA 1. Note that this is the first time that H$\alpha$ has been found to be tracing an outflow from a BD. Also note that both lobes of the flow are seen in H$\alpha$ while in the FELs only blue-shifted emission is detected. This has been observed in CTTSs and may be explained by the presence of a ``dust hole" in the disk, close to the central star (14, 15).  As the H$\alpha$ is tracing the outflow much closer to the BD than the FELs, its red-shifted component can be seen through the dust hole in the disk. The FELs originate much further from the star where the red-shifted emission remains obscured by the disk. This result can be taken as further evidence that BD disks evolve in a similar fashion to T Tauri disks and may at some stage harbour planets.  Figure 1 shows the results of the spectro-astrometric analysis of the [SII]$\lambda$6731 and H$\alpha$ lines.}

\end{itemize}

As stated, the long-term aim of this project is to make a comprehensive comparison between protostellar outflows/jets and those driven by BDs. So far results have revealed that optical outflows driven by accreting BDs are common. Notable T Tauri-like properties of the BD outflows include, the predominance of blue-shifted emission (indicating the presence of an accretion disk) and the fact that the measured offsets in the BD FELs lie within the range estimated from BD if BD outflows scale down from T Tauri-jets (16). In order to truly test whether the same the T Tauri outflow mechanism is operating in the sub-stellar regime, images of BD outflows are needed to investigate collimation. In addition, comparisons between mass outflow and infall rates must be made.  

%
%
%


\end{document}